\documentclass[preprint]{aastex}

\shorttitle{RWI in Disk/Planet System}

\shortauthors{Ou, Ji, Liu, Peng}

\begin{document}


\title{{Disk-Planet Interaction Simulations: (I) Baroclinic Generation of Vortensity and
          Non-Axisymmetric  Rossby-Wave-Instability}}


\author{Shangli Ou\altaffilmark{1}, Jianghui
JI\altaffilmark{2,3,4}, Lin  Liu\altaffilmark{5}, and Xiaomeng Peng\altaffilmark{6}}

\altaffiltext{1}{High Performance Computing, Center for Computation and Technology / 
Information Technology Services,
Louisiana State University, Baton Rouge, LA, 70803;ou@cct.lsu.edu}

\altaffiltext{2}{Purple  Mountain  Observatory , Chinese  Academy
of  Sciences,  Nanjing  210008,China;jijh@pmo.ac.cn}

\altaffiltext{3}{National Astronomical Observatory, Chinese
Academy of Sciences, Beijing 100012, China}

\altaffiltext{4}{Department of Terrestrial Magnetism, Carnegie
Institute of Washington, 5241 Broad Branch Road NW, Washington, DC
20015-1305}

\altaffiltext{5}{Department of Astronomy,  Nanjing University,
Nanjing  210093, China}

\altaffiltext{6}{Department of Physics \& Astronomy,  
Louisiana State University, Baton Rouge, LA, 70803}


\begin{abstract}

We use a multi-dimensional hydrodynamics code to study
the gravitational interaction between an embedded planet and a protoplanetary disk
with emphasis on the generation of vortensity (potential vorticity) through a Baroclinic Instability and 
subsequent development of Rossby-Wave-Instability (RWI).
It is found that the generation of potential vorticity
is very common and effective in non-barotropic disks
through the Baroclinic Instability, especially within the coorbital region.
Our results also complement previous studies by \citet{KLL03}
that non-axisymmetric RWIs are likely to develop
at local minima of potential vorticity distribution
that are generated by the interaction between a planet and a inviscid barotropic disk.
This second instability appears to be very common and robust, regardless of
the equation of state, initial density distribution, and rotational law of the disk.
The development of RWIs results in non-axisymmetric density blobs,
which exert stronger torques onto the planet when they travel
in the vicinity of the planet.
As a result of that, large amplitude oscillations are introduced 
to the time behavior of the total torque acted on the planet by the disk.
In our current simulations, RWIs do not change the overall picture of inward
orbital migration but bring in a non-monotonic behavior to the migration speed.
As a side effect, RWIs also introduce interesting structures into the disk.
These structures may help the formation of Earth-like planets in the Habitable Zone
or Hot Earths interior to a close-in giant planet.

\end{abstract}


\section{Introduction}
Although theorists did not realize its relevance to the origin of
planet when the study on the gravitational interaction between a
protoplanetary disk and an embedded planet just began in late
1970s \citep{GT78}, it is now one of the key ingredients of
current standard theory of planet formation.
According to standard core-accretion theory \citep{Saf69,Lis93}, the
formation of planets in circumstellar disks around T Tauri stars
consists of the formation of planetesimals via
collisions/coalitions of dust grains in the early stage and then
gravitationally accretion after they accumulate enough mass. The
life time of this T Tauri star phase is estimated to be short
($\lesssim 10^7$ years). In order for cores of protoplanets
with masses comparable to that of Jupiter to accumulate enough
mass in the T Tauri stage, it is thought that their cores have to
form outside the so-called ``ice line", where the distance to the
central star is large \citep[typically beyond $\sim 4$ AU, see][]{Ida04}
so that the temperature is low enough to allow the
condensation of gas materials to solid ice. These additional solid
dusts helps to increase the dust coagulation speed and shorten the
time needed to form cores of protoplanets.

Since the discovery of the first Jupiter-mass planet \citep{May95}
orbiting the solar-type star 51 Peg, more than 200 extrasolar planets have been
discovered around the nearby stars within 200 pc
\citep[] [The Extrasolar Planets Encyclopaedia]{But06}.
The diversity exhibited by
these planetary systems shows that the masses of these planets
range from Jupiter-mass to Neptune-mass. The most prominent
characteristics is that a large number of the host stars are
surrounded by the so-called ``hot-Jupiters" and close-in
super-Earths. Around $80\%$ of the extrasolar planets are in
orbits with semi-major axes in the range $0.01 \lesssim a \lesssim
2.5$ AU, and $\sim 25\%$ of the total population are
short-period planets\footnote{see http://exoplanet.eu/} with $a
\lesssim 0.1$ AU. This has brought one of the most interesting
puzzles to theorists: if protoplanets had formed in a disk region
beyond $\sim 4$ AU from the central star, how did the observed
extrasolar planets end up with orbits that are so close to their
host stars? If the standard theory for the formation of
protoplanetary cores holds (Pollack et al. 1996; Ida \& Lin 2004),
then the giant planets (like hot-Jupiters) must have undergone
inward orbital migration to their current locations
\citep{Lin96}.

Several mechanisms have been proposed as the drive for the
migration. Some authors \citep{Wei96, RF96, For99}
suggested that three-body interaction
between a central star and two or more planets can lead to the
ejection of one of the planets, which takes away energy from the
system and leaves the third object on a smaller and eccentric
orbit, which is finally circularized through tidal interaction
between the planet and its host star. \citet{mur98} suggested that
resonance interactions between a planet and a disk of planetesimals can
also lead to inward orbital migration of the planet. Another plausible
explanation is provided by the theory of the gravitational
interaction between a protoplanetary disk and an embedded planet,
which was formulated, as mentioned in the beginning of this
article, more than a decade before the discovery of extrasolar
planets.

In the third mechanism, a disk interacts with an embedded planet
through their mutual gravitational forces. The planet causes the
formation of spiral waves inside the disk at the Lindblad
resonances; as a result of the density asymmetry induced by spiral
waves, the inner disk exerts a positive gravitational torque onto
the planet and the outer disk exerts a negative gravitational
torque onto the planet. According to analytical analysis,
the overall torque is generally negative and, hence,
forces the planet to migrate inward.
\citet{War97} pointed out that two major kinds of migration exist
 based on the mass of the embedded protoplanet.
In the so called Type I migration, the planet's mass is small and
the response of the disk is linear; the migration speed is very
fast so that the migration timescale is as short as $\sim
10^{4}$ yr for a planetary core of $10M_{\oplus}$ in a sufficiently
viscous disk. In Type II migration, the protoplanet has enough
mass to open a gap inside the disk, and migrates on a much longer viscous
timescale.
Besides classical analytical analysis,
many groups have studied the nonlinear evolution of
a disk-planet system using numerical multi-dimensional hydrodynamics
\citep[][and references there in]{Kle99,NPMK00, Pap01, NW03, bate03}.
Recently, \citet{deval06} also carried out a collaborative multi-code comparison
research.
These numerical simulations showed that the nonlinear evolution of
the orbital migration of a planet inside a disk agrees with linear analysis
in a qualitative manner.

However, nature is always not as simple as human beings have naively thought.
Unexpected phenomena often arose in the non-linear regime.
For example, \citet{mas01} and \citet{mas03} suggested that the corotation torque
originated from the coorbital region
may play a very different role from that of Lindblad torques;
this leads to a third kind of migration often
referred to as Type III migration or runaway migration,
in which the migration happens on a timescale as short as a few tens of orbits
and can be directed outward in some cases.
\cite{klahr03} described a Baroclinic Instability in non-barotropic disks
that may contribute to vorticity and global trubulences;
They argued that strong vorticities may contribute to the
rapid formation of Jupiter-size gas planet \citep{klahr06}.
\citet{KLL03} and \citet{LL05} showed that the so-called ``Rossby-wave instabilities"
may develop at the local minima of potential vorticity (PV), or vortensity
(defined as the ratio between local vorticity and surface density),
in an inviscid disk with initially uniform vortensity distribution.

In simulations of \citet{LL05},
these non-axisymmetric instabilities lead to the formation of vorticies and
density blobs, which exert stronger torque onto the planet when they travel
around its vicinity and bring large oscillations to the total torque
acted on the planet.
They further argued that this mechanism may be possible to change the direction
of the migration.
Non-axisymmetric Rossby-wave instabilities are also relevant to
the evolution of a single disk \citep{LLCN99, LFLC00} and
stellar models with strong differential rotation \citep{OT06}.
The time scale for RWIs to fully develop is generally on tens of dynamical time
of a disk ($\sim 100$ orbits), hence, they can not be ignored in inviscid or
low viscosity cases.
Therefore, it worths more labor to study their comprehensive role on planetary orbital migration
and the robustness of this mechanism.

In this paper, we present results from simulations of disk-planet interaction.
Our purpose is two-folded:
first we wish to provide further comparisons to results presented in
the collaborative comparison work of \citet{deval06};
on the other hand, we focus more on the development of non-axisymmetric RWIs in
disk models with non-barotropic equation of state (EOS) and
non-uniform initial vortensity distribution, 
as in contrast to disk models presented in \citet{KLL03} and \citet{LL05},
which are isothermal disks with uniform initial vortensity everywhere.
In particular, we find out that the PV generation 
is more effective and common in disk models with non-barotropic EOS.
Our results, together with those presented by \citet{LL05},
suggest that the development of RWIs is very robust in disk-planet
systems with different mass distribution, rotational laws and EOSs.
We also address the impact of RWIs on planet orbital migration
and formation of close-in super Earths near a giant planet.
In section 2, basic equations, methods and initial models are described;
results from test runs are given in section 3;
section 4 concentrates on the development of RWIs under certain circumstances;
we close with conclusions and discussions in section 5.

\section{Basic Equations, Methods and Initial Setup}
To study the interaction between a disk and an embedded planet
requires coupling hydrodynamics and orbital dynamics together.
Here, we follow \citet{NPMK00} and many previous investigations
 to reduce the problem to a two-dimensional (2D) one
since the disk is considered to be very thin.
Three dimensional (3D) investigations will be postponed to future.
The fluid motion inside the disk is described by the vertically integrated
 continuity equation (\ref{conteq}), radial and azimuthal components of
the Navier-Stokes equation (\ref{NS_eq_r}) and (\ref{NS_eq_p}),
\begin{eqnarray}
   \frac{\partial \Sigma} {\partial t} + \nabla \cdot (\Sigma \vec{v}) &=& 0 \label{conteq}\\
   \frac{\partial(\Sigma v_r)} {\partial t} + \nabla \cdot (\Sigma v_r \vec{v})
      &=& \frac{\Sigma v^2_{\phi}} {r} - \frac{\partial P}{\partial r}
         - \Sigma \frac{\partial \Phi}{\partial r} + f_r  \label{NS_eq_r}\\
   \frac{\partial(\Sigma v_{\phi})} {\partial t} + \nabla \cdot (\Sigma v_{\phi} \vec{v})
      &=& -\frac{\Sigma v_r v_{\phi}} {r} - \frac{1}{r} \frac{\partial P}{\partial \phi}
         - \frac{\Sigma}{r} \frac{\partial \Phi}{\partial \phi} + f_{\phi} \label{NS_eq_p} \,,
\end{eqnarray}
where $\Sigma$ is disk surface density, $\vec{v}$ is two fluid
velocities, $P$ is vertically integrated pressure, $f_r$ and $f_{\phi}$ are two
components of viscous forces, and $\Phi$ is the gravitational
potential felt by fluid elements. 
Details regarding viscous terms can be found in \citet{NPMK00}. 
The EOS of the disk fluid is considered as locally isothermal \citep{NPMK00} 
as given in below
\begin{equation}
P = c_s^2 \Sigma \,,
\end{equation}
where the local isothermal sound speed is $c_s = \frac{H}{r} \sqrt{GM_*/r}$
with disk aspect ratio $H/r=0.05$. 
(As will be discussed in details later, this radial variation of sound speed
generates vorticity wherever an azimuthal density gradient exists.)

We further simplify our study to
non-self-gravitational systems, in which the self-gravity of the
fluid is not taken into account for the fluid motion; hence, $\Phi
= \Phi_* + \Phi_p$, where $\Phi_*$ is the potential field of the
central star and $\Phi_p$ is the potential field of the planet,
which is given by $\Phi_p=- M_p /\sqrt{r^2+\epsilon^2}$, where
$M_p$ is the planet mass and
$\epsilon$ is taken to be 0.2 of the Roche Lobe of the planet. 
The initial disk model has Keplerian rotational profile and uniform
density, which results in an initial radial vortensity profile
$\xi(r)$ that is proportional to $r^{-\frac{3}{2}}$ (vortensity is
defined as the ratio between local vorticity and surface density).
The value of density and viscosity are chosen to follow those
specified in \citet{deval06}.
The disk and the non-rotating coordinate system
are centered at the central star instead of the center of
mass of the system, which locates slightly off the central star.
To handle the hydrodynamics part, we adopted a legacy code
developed by the astrophysical group at Louisiana State
University to study star formation \citep{T80, WT88},
mass transferring binary stars \citep{MTF02},
and rotating instabilities in neutron stars\citep{OT06}.
The code is explicit and 2nd order in both space and time. It splits the
source term and advection term in a manner similar to Zeus
\citep{SN92}. Other features implemented include Van Leer upwind
scheme, artificial viscosity to handle shock, and, staggered
cylindrical grids. The code is originally three-dimensional,
but adapted to 2D in this work.
At the boundary of our computational grids, 
mass is allowed to flow off the grids but no inflow is allowed.
We also implemented the wave-killing boundary condition
specified in \cite{deval06} for comparison purpose.

The equation of motion for the planet is:
\begin{eqnarray}
   \frac{d^2 \vec{r}_p} {dt^2} = -\frac{G(M_*+M_p)}{r_p^3} \vec{r}_p -  
       G \int \frac{\Sigma(\vec{r^\prime})}{(|\vec{r}_p - \vec{r^\prime} |^2 + \epsilon^2)^\frac{3}{2}} (\vec{r}_p - \vec{r^\prime} ) r^\prime dr^\prime d\phi 
\end{eqnarray}
where the first term on the right is derived from the relative motion
of the planet to the central star,
the second term is the gravitational force exerted on the planet 
by the disk.
For comparisons with previous investigations \citep{LL05,deval06},
we turned off the disk's attraction in many of our simulations
so that the planet stays on a fixed circular orbit at 1 $AU$;
for certain runs, the disk potential was turned on to allow
the planet to move so that we can assess RWI's role on the migration.
A 4th order Runge-Kutta integrator is used to integrate forward
the equation of motion of the planet during each Courant timestep
as required by the hydro part.
To compute the torque acted on the planet, we followed the specification in \citet{deval06},
\begin{equation}
   T_z = GM_p \int \Sigma \vec{r}_p \times \frac{ (\vec{r^\prime}-\vec{r}_p )}
        { ( |\vec{r^\prime}-\vec{r}_p |^2 + \epsilon^2)^\frac{3}{2}}  r^\prime dr^\prime d\phi \,,
\end{equation}
torques from inside and outside the Roche-lobe of the planet are computed separately
for both inner disk and outer disk.
In the following sections, we show time evolution of the total torque including 
contributions from within and outside the Roche-lobe, 
the temporal behavior of the total torque does not change qualitatively when torques from within
the Roche-lobe are excluded.

In equations of motion for both the planet and disk, we neglected the indirect terms
due to the acceleration to our coordinate system by the disk
(the planet's acceleration to the coordinate system is neglected 
for the disk evolution as well).
Since the difference between the center of the mass and
our coordinate origin is fairly small, we expect the effect of indirect terms
won't be significant to change our results qualitatively;
hence we dropped these terms from our equations to save computing time.
(The evaluation of these indirect terms involves integrals over
the mass distribution and is therefore relatively expensive.)
Finally, the accretion onto the planet is neglected as well.

The units adopted are the following: the gravitational constant
$G=1$, length unit is 1 $AU$, and $M_*+M_p=1$. These setups help
us to have a direct comparison between our results and those
presented in \citet{deval06}.
A Jupiter mass is defined as $10^{-3}$ and a Neptune mass is
defined as $10^{-4}$. The planet's mass is turned on gradually in
the first 5 orbits.
As shown in \citet{LL05},
very high resolution is required to resolve RWIs.
This limits our ability to push simulations to much longer time scale.
In general, we only advance our simulations to $ \sim 200$ orbits,
which is sufficient for RWIs to fully develop.

\section{Test Runs for Jupiter mass planet}

In this section, we present results for interaction between a disk
and a Jupiter mass planet on a fixed orbit as a calibration to our code.
According to previous analytical analysis and numerical nonlinear studies \citep{LP86,TBGLGB98,NPMK00},
a Jupiter mass planet
is capable of clearing most of the materials around its orbit and
opening a fairly wide and deep gap in a viscous/inviscid disk.
In order to test whether our code can simulate this correctly,
we have run this configuration for both viscous and inviscid disks
on grids with two different resolutions,
$128\times384$ and $256 \times 512$, to check for convergence.
To avoid lengthy descriptions, we only show results for a viscous disk.

\begin{figure}[]
\epsscale{1.0} \plotone{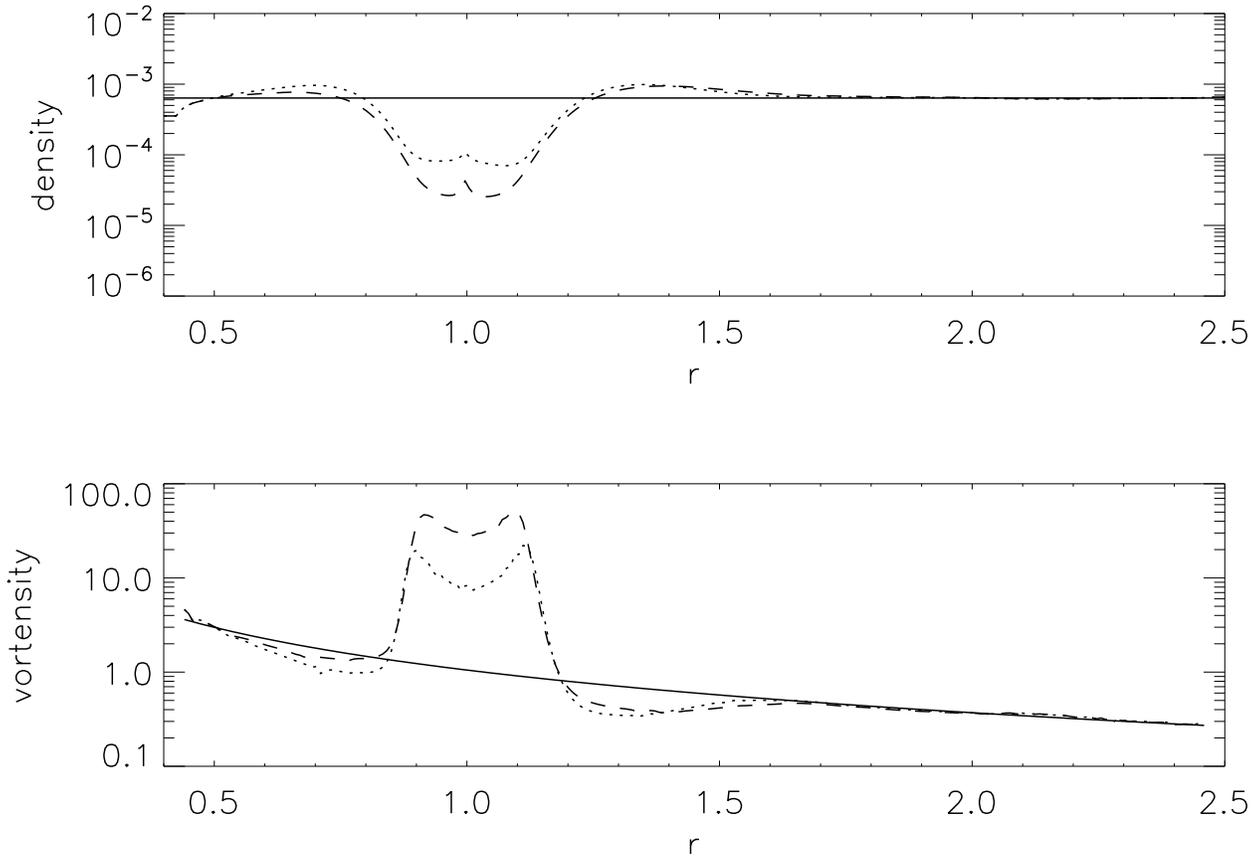}
\caption{The
top and bottom panels show, respectively, $\Sigma(R)$ and $\xi(R)$
of a viscous disk with a Jupiter mass planet averaged azimuthally  at $t
\approx 0$ (solid line), $100$ (dotted line), and $200$ (dashed
line) orbits for the higher resolution run ($256 \times 512$).
\label{fig:rho_vort_jup} }
\end{figure}

\begin{figure}[]
\epsscale{1.0} \plotone{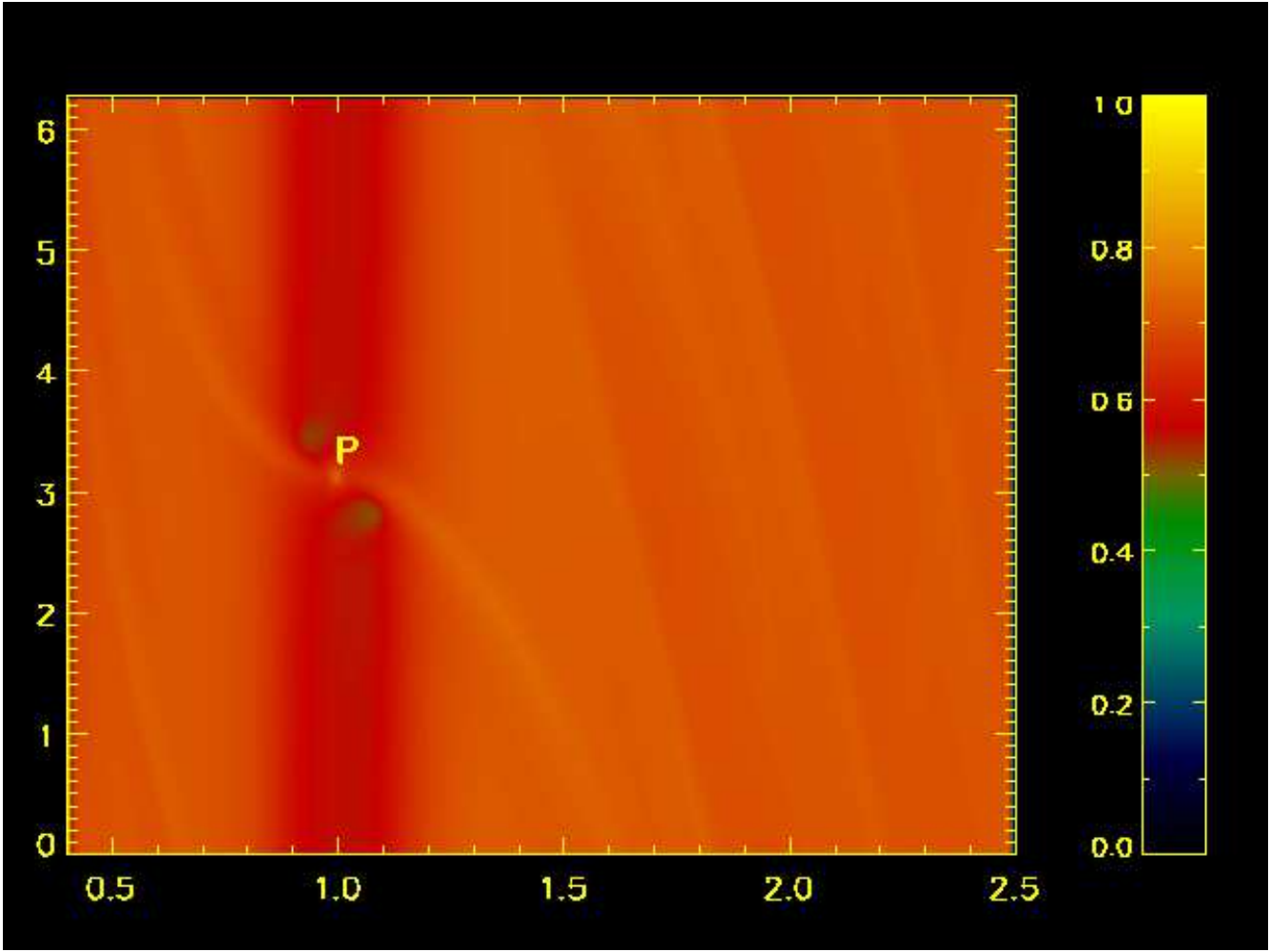} 
\caption{Logarithmic
color map of surface density distribution for a viscous disk with
a Jupiter mass planet at $t\approx 200$ orbits for the higher
resolution run ($256 \times 512$). The horizontal axis is the
radial axis ranging from 0.4 to 2.5, and the vertical axis is the
azimuthal direction spanning over $2\pi$ range with the planet
shifted to around the middle.
A letter ``P" is labeled next to the location of the planet.
The color bar represents relative rather than absolute values.
}
\label{fig:isoden_jup}
\end{figure}

Figure \ref{fig:rho_vort_jup} displays radial profiles of surface density
$\Sigma$ and vortensity $\xi$ averaged over the azimuthal
direction at $t \approx 0, 100, 200$ orbits for the higher
resolution run. At the final age, a wide and deep gap has been
opened from the initial flat density profile, the contrast between
the lowest and highest density spans over around two orders of
magnitude. Inside the gap where the surface density is extremely
low, the vortensity has increased to a very high level. (Notice
that the vortensity is normalized to its initial value at $r=1$.)
Two vortensity minima formed around the inner and outer edge of
the gap, where RWIs are expected to grow. But since surface
density inside the gap is extremely low, we don't expect it to
have significant effect on the evolution.
(We note that the existence of viscosity also affects
the development of RWIs.)
 To compare with plots
presented in \citet{deval06}, 
a logarithmic color map of surface density distribution of the disk
at $t \approx 200$ orbits
is shown in Figure \ref{fig:isoden_jup}. 
It is observed that spiral arms originating from
the planet forms in both the inner and outer disks. These two
spiral arms are well-known Lindblad waves caused by the potential
of the planet and, in turn, exert a negative total torque onto the
planet as predicted by classical theory \citep{GT78, LP80}. The
structures shown in this plot agree qualitatively with those shown
in \citet[][see their Fig. 10]{deval06}.

\begin{figure}[]
\epsscale{1.0} \plotone{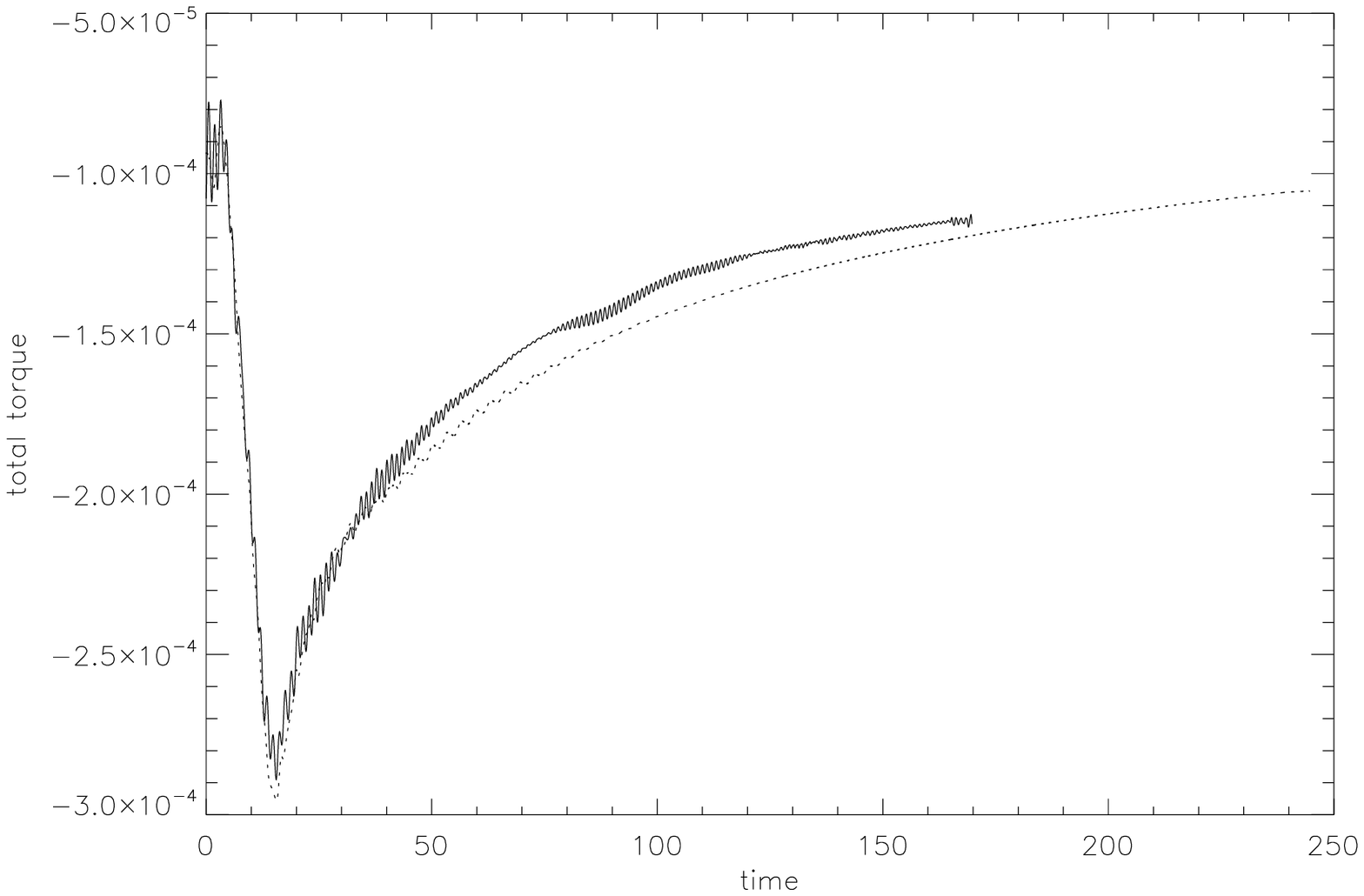}
\caption{Time evolutions of the total torque (per unit mass) exerted on a Jupiter mass planet
by the disk for lower (solid line) and higher (dotted line) resolutions}
\label{fig:torq_jup}
\end{figure}

Figure \ref{fig:torq_jup} illustrates the evolution of the total torque
(per unit mass) exerted on the planet by the disk for both lower
and higher resolution runs. The data have been smoothed using a
moving box method over 10 orbit period \citep{NPMK00}. Their
behaviors agree closely with each other, which suggests that we
have achieved convergence on these results. As expected, the total
torque has a slightly negative value and will drive the planet to
migrate inward gradually. The magnitude of the torque falls in the
same range of those presented in a recent collaborative
comparison study \citep[][see their Table 6]{deval06}.
We also monitored the mass in the computational grids,
there are around 93\% of the initial mass left
at the end of the higher resolution run.

In general, our test runs for Jupiter mass planet agree with previous investigations in a qualitative manner.
In the following, we will switch to the situation of a Neptune-mass planet embedded in a protoplanetary disk
and study the development of RWIs in such configurations.

\section{High Resolution runs for RWI in protoplanetary disk with a Neptune-mass planet }

\subsection{Results for simulations with a Neptune mass planet on a fixed orbit }
In this subsection, we present results from a series of
simulations in which a Neptune-mass planet is on a fixed circular orbit.
Previous investigations \citep{NPMK00,deval06} have shown that a
Neptune-mass planet can only creates a shallower and narrower
gap inside a disk due to its smaller mass. \citet{KLL03} and \citet{LL05}
further suggested that, in an inviscid disk with a Neptunian mass
planet, radial vortensity minima are important because they may
trigger RWIs to develop. As a result of RWI, non-axisymmetric
density blobs form and orbit around the
central star; the torque exerted on the planet by the disk can
undergo large amplitude oscillations as these density blobs pass
by the vicinity of the planet and apply gravitational
perturbations to it. As suggested by \citet{LL05}, this
mechanism may potentially slow down the inward migration of a
planet and, in some extremely cases, even change the sign of the
total torque exerted on the planet and, thus, cause it to migrate
outward.

Since the disk models studied in \citet{KLL03} and \citet{LL05} have 
an isothermal EOS and an initial
flat vortensity profile, we wish to investigate if the same
mechanism will also work in disks with non-barotropic EOS and  non-uniform vortensity.
With this in mind, we adopted Keplerian disk models similar to
those of \citet{deval06}, which has a locally isothermal EOS and an initial uniform surface
density and, hence, an initial monotonically decreasing vortensity
profile as proceeding outward radially. We carried out simulations
of this configuration on grids with resolutions of $128\times
384$, $400\times800$, and $400\times1600$; 
a simulation with a resolution of $800 \times 3200$ was also performed for $\sim 70$ orbits
to check for the convergence of vortensity profile. The viscosity is turned
off in these simulations.

In general, our simulations show that the evolution of the system consists of
two stages: the first stage is the formation of two spiral arms which exert
a negative torque to the planet; the second stage is the development of
RWIs in valleys (minima) of the radial vortensity profile
some time after the formation of spiral arms.
However, fine structures are only resolved accurately in higher resolution runs,
mainly because interactions happen mostly around the co-orbital region
\citep{LL05}.
In the following, we show detailed results from a run with a resolution of $400\times1600$.

\begin{figure}[]
\epsscale{1.0} \plotone{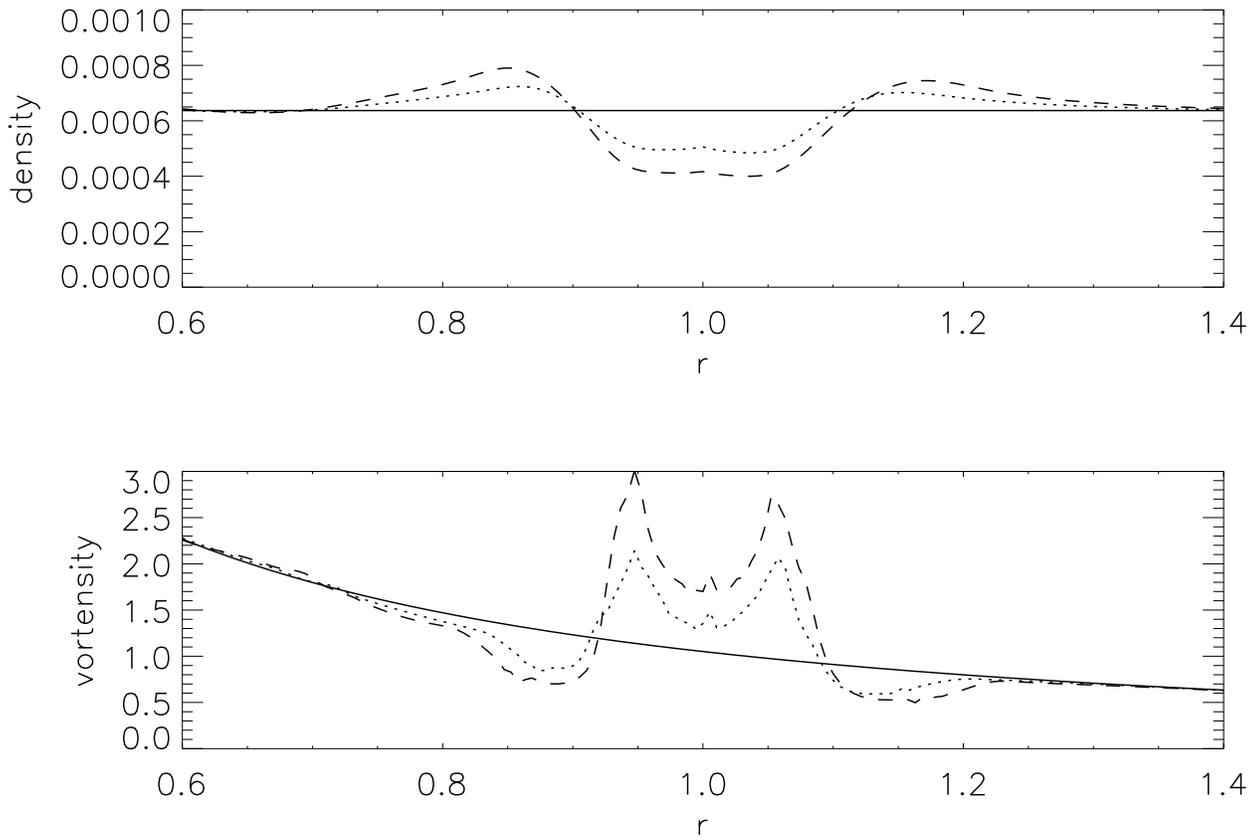}
\caption{The top and bottom panels show, respectively, $\Sigma(R)$
and $\xi(R)$ of the disk with a Neptune mass planet averaged azimuthally
at $t \approx 0$ (solid line), $60$ (dotted line), and $120$ (dashed line) orbits.
\label{fig:rho_vort_nep} }
\end{figure}

Figure \ref{fig:rho_vort_nep} displays radial profiles of
$\Sigma$ and $\xi$ averaged over the azimuthal direction at $t
\approx 0, 60, 120$ orbits. In order to give a better view of fine
structures around the co-orbital region, only regions from $r=0.6$
to $r=1.4$ are shown in the plot. Compared to the case of a
Jupiter mass planet, a Neptune mass planet only opens a relatively
shallower and narrower density gap. The contrast between the lowest
and highest density is around a factor of two. Since the
surface density inside the gap is not quite low, the vortensity
there does not increase to much high level. (Notice that the
vortensity is normalized to its initial value at $r=1$.) However,
three vortensity valleys formed across this radial region, with
one centered around the planet's orbit. Because the density inside
the gap is still significant, RWIs resulted from these vortensity
minima will have important effect on the evolution of the system
\citep{LL05}.

\begin{figure}[]
\epsscale{1.0} \plotone{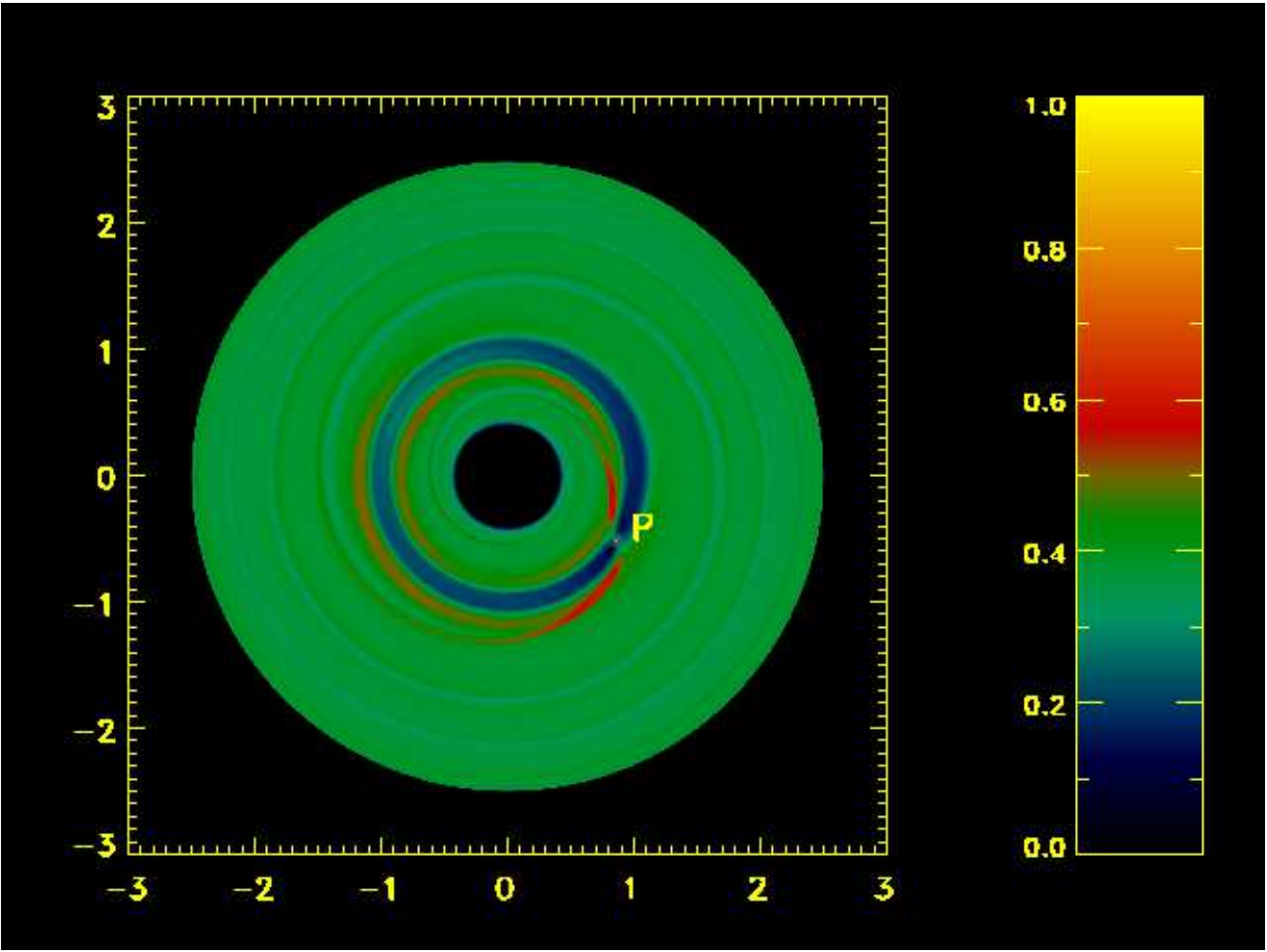}
\caption{The same as Fig. \ref{fig:isoden_jup} but in linear scale
for an inviscid disk with a Neptune mass planet at $t\approx 150$ orbits.
 A letter ``P" is labeled next to the location of the planet.
The color bar represents relative rather than absolute values.}
\label{fig:isoden_nep}
\end{figure}

A linear color map of surface
density distribution on a polar plot \citep{LL05} at  $t \approx
150$ orbits is illustrated in Figure \ref{fig:isoden_nep}. 
Focusing on the formation of high density
areas (red/brown regions), we observe not only red
Lindblad spiral arms, but also, other non-axisymmetric high
density structures at different locations: the inner edge of the
outer disk (brown arc-shaped region centered around 7 oclock) , the outer edge
of the inner disk (red/brown arc-shaped region around its edge);
interestingly, the density inside the gap is no longer
axisymmetric any more,
as suggested by the light blue region around 10 oclock.
The radial locations of these non-axisymmetric
structures match exactly with the local minima of the vortensity
distribution (see bottom portion of Figure \ref{fig:rho_vort_nep})
 as predicted by linear analysis \citep{LFLC00}.

\begin{figure}[]
\epsscale{1.0} \plotone{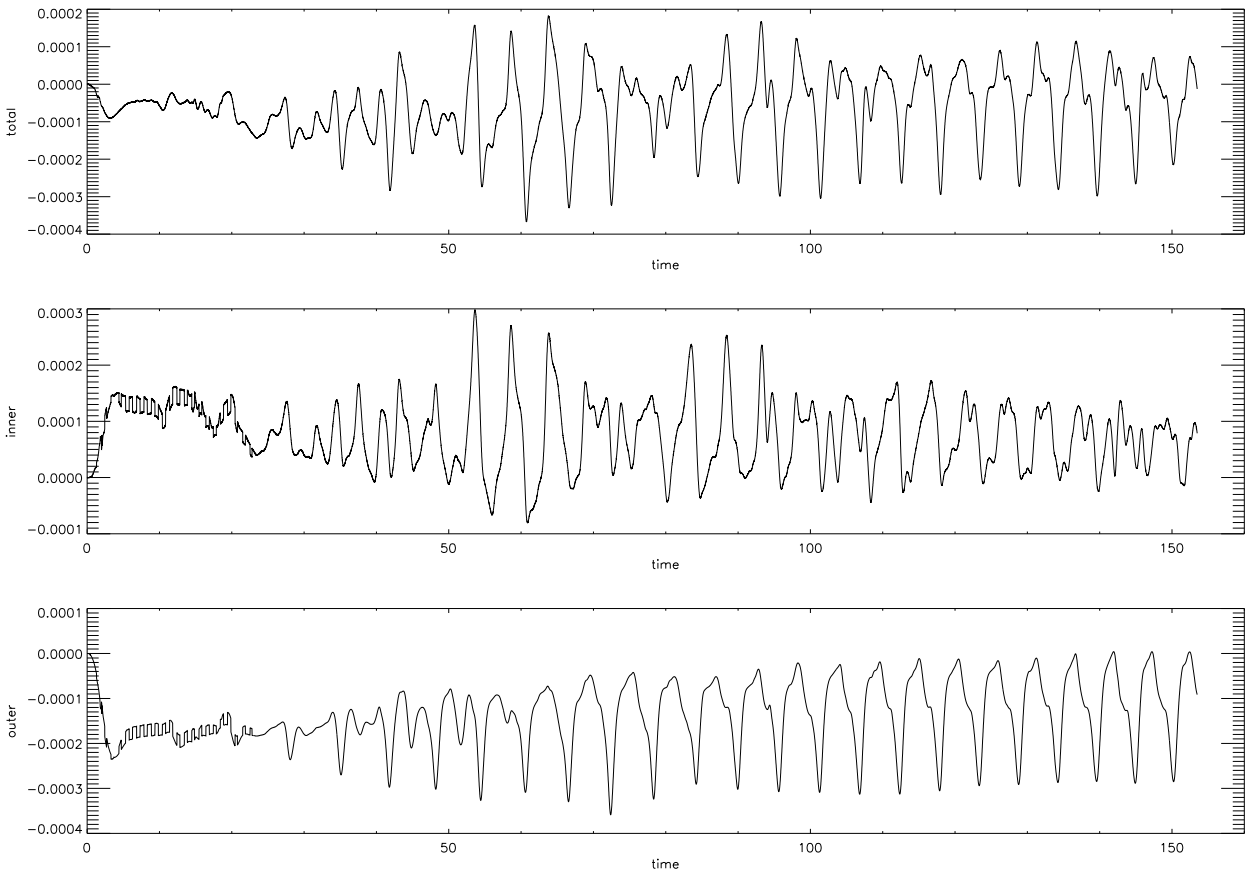}
\caption{Time evolutions of total torque (top), torques from the
inner (middle) and outer (bottom) disks exerted on a Neptune mass planet
for the highest resolution run ($400 \times 1600$). Raw data is shown.}
\label{fig:torq_nep}
\end{figure}

Figure \ref{fig:torq_nep} shows the time evolution of the total torque on the planet,
 the torque from the inner and outer disk.
The behavior of the total torque is totally different from that in the Jupiter mass case:
very large amplitude oscillations present through most of the simulation.
As pointed out by \citet{LL05}, these oscillations are caused by
the gravitational pull from those orbiting density blobs;
each time when they pass the vicinity of the planet,
they exert a stronger torque onto the planet,
the sign of their torques depends on their relative position to the planet.
The existence of these oscillations complicates the classical picture of
the gravitational torque exerted on the planet by the disk.

\begin{figure}[]
\epsscale{1.0} \plotone{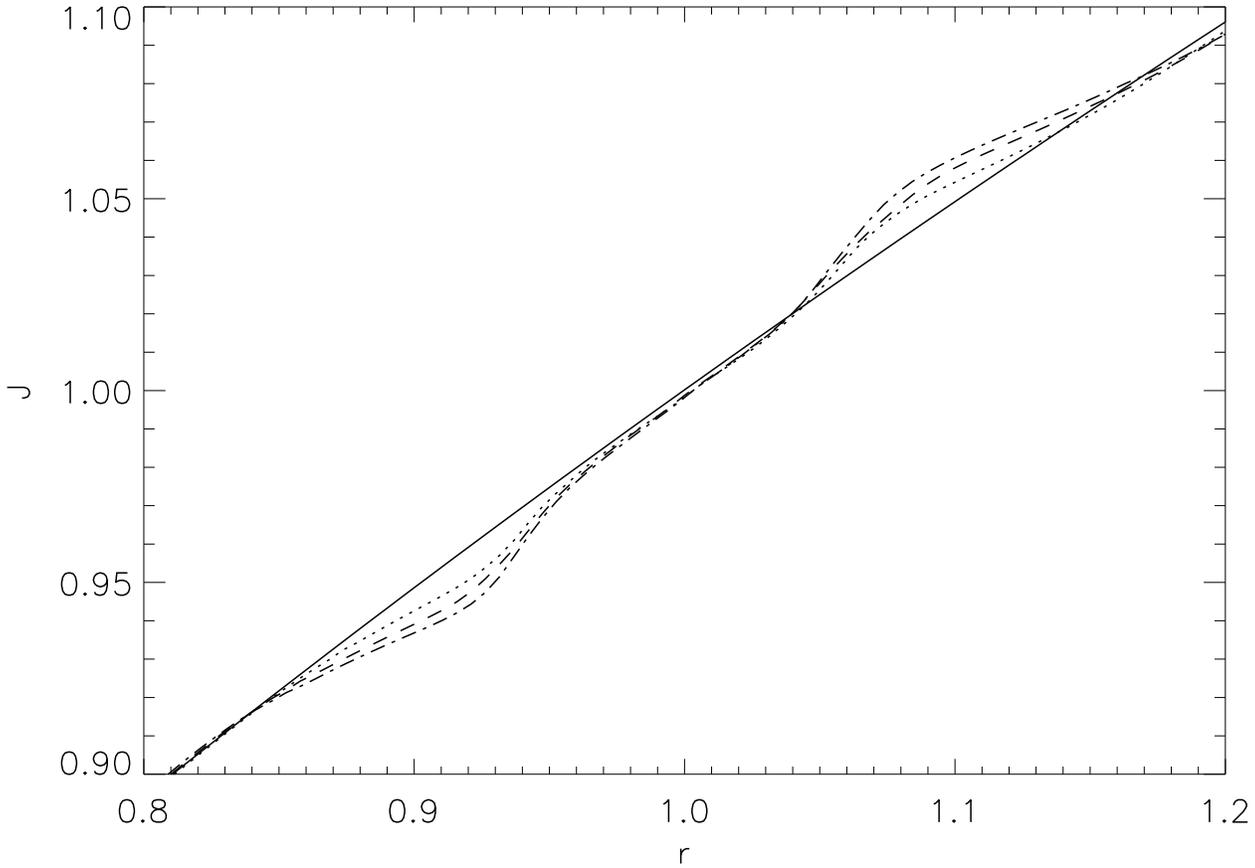} 
\caption{Radial profiles of
specific angular momentum around the co-orbital region of the disk
with a Neptune mass planet at different times. The solid line
denotes for initial Keplerian profile, dotted, dashed, and
dotted-dashed lines denote for profiles at later times. Data have
been normalized to the initial value at $r=1$. As evolutions goes
on, angular momentum is being removed in the inner region but
added to the exterior region due to the interaction between the
disk and the planet.} \label{fig:jn_nep}
\end{figure}

Figure \ref{fig:jn_nep} exhibits the radial profile of the specific
angular momentum $J$ inside the disk around the co-orbital region
at different time. The solid line denotes for initial Keplerian
profile, dotted, dashed, and dotted-dashed lines denote for
profiles at later times. As the evolution goes on, the inner disk
exerts a positive torque to the planet and, hence, is giving away
angular momentum to the planet; whereas, the outer disk exerts a
negative angular momentum to the disk, thus, is gaining
angular momentum from the planet; therefore, the angular momentum
is being transferred from the inner disk to the outer disk due to
the interaction between the planet and the disk. This mechanism
results in a dip in the inner disk's $J$ profile but a bump in the
outer disk's $J$ profile. As a result, an S-shaped structure
formed in the vicinity of the planet's fixed orbit.
Again, for comparison to previous studies,
there are around 98\% masses left in the grid
at the end of the highest resolution run.

In summary, our results of the development of RWIs in 
an inviscid disk with a Neptune-mass planet on a fixed orbit 
are in agreement with those of \citet{LL05};
however, they differ in some key physical quantities.
In particular, the PV maxima in \citet[][see their Figure 2 and 9]{LL05} are located at radii outside 
the coorbital region where shocks present;
whereas, PV maxima in our simulations reside within the coorbital region
around $0.95 < r< 1.05$, where shocks are not expected to occur.
This discrepancy raises some concerns. 
Because based on the evolution equation of vorticity $\vec{\zeta}$:
\begin{equation}
   \frac{d \vec{\zeta}}{dt} = \frac{\nabla \Sigma \times \nabla P}{\Sigma^2}
\end{equation}
the vorticity can only be generated at regions
where pressure gradient and density gradient do not align with each other;
such generation of vorticity normally happens around shocks in barotropic disks as illustrated in \citet{LL05},
which contributes to vortensity maxima outside the coorbital region.

\begin{figure}[]
\epsscale{1.0} \plotone{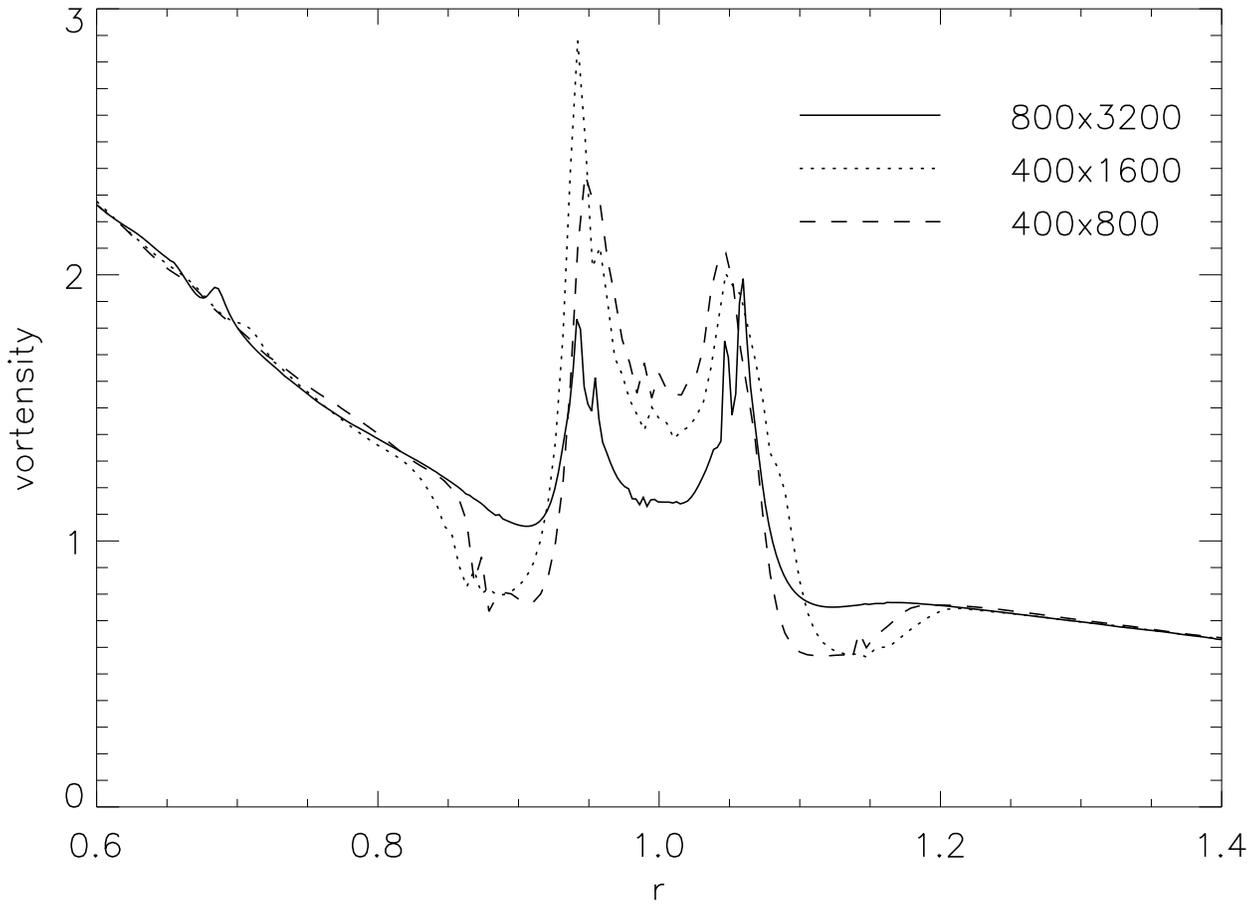}
\caption{Vortensity profiles around the coorbital regions for runs with different resolutions
at $t \approx 70$ orbits.
All of them show maxima within the coorbital region.
} \label{fig:vorten_conv}
\end{figure}

\begin{figure}[]
\epsscale{1.0} \plotone{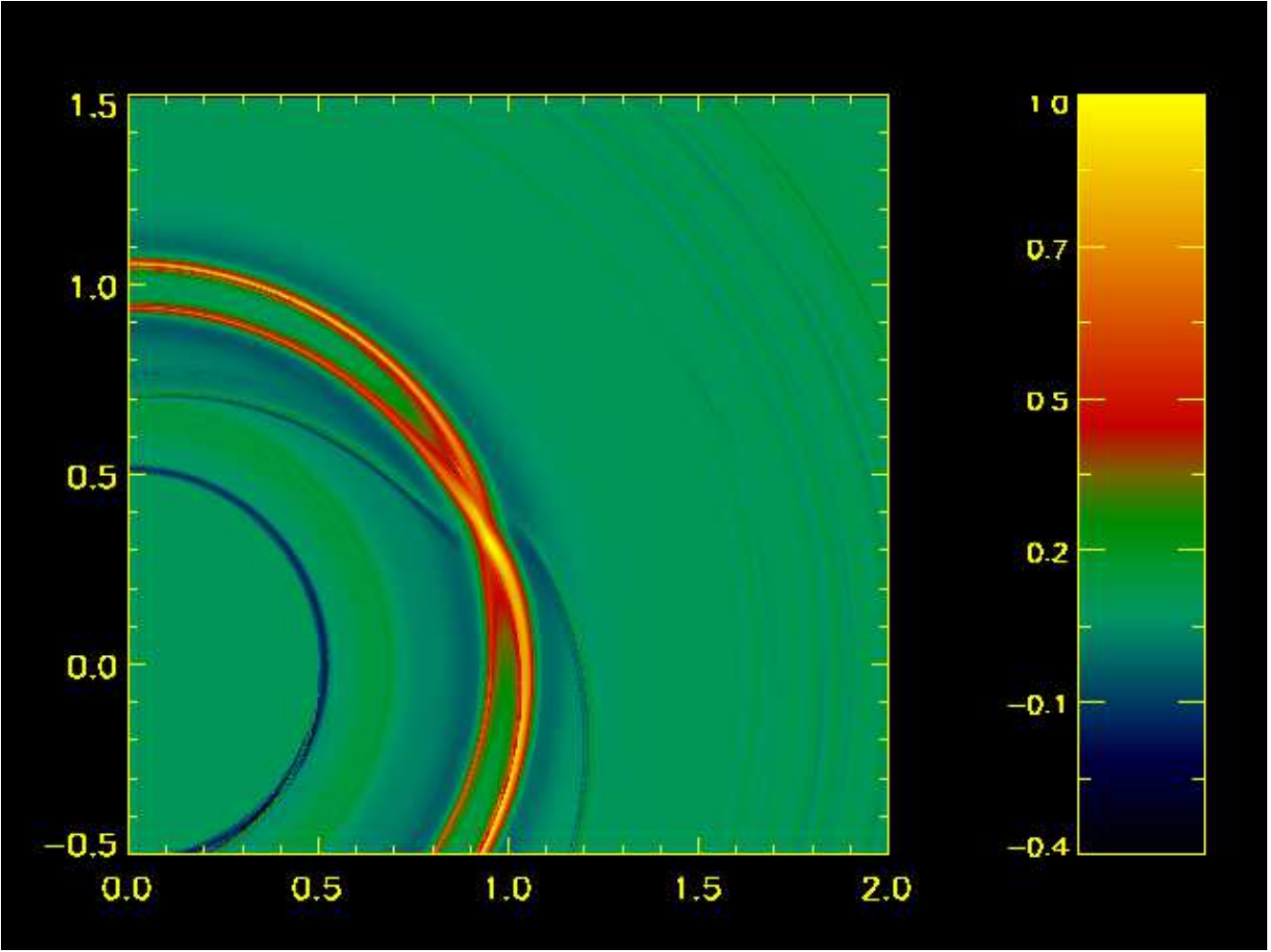}
\caption{Distribution of vortensity increment around the planet at $t \approx 70 $ orbits
for the run with resolution $800 \times 3200$.
(The initial backrougnd vortensity is subtracted.)
Shocks are well resolved. Vortensity is generated within the coorbital region,
especially in the Roche lobe, where no shock exists. 
The azimuthal density gradient within the same region
 acts as a source term of vortencity generation (see Figure \ref{fig:drho_HR}).
Note that the vortensity increment within the Roche-lobe is always positive,
which contradicts our expectation. We suespect this may be resulted from 
vortensity mixture in this region.
} \label{fig:vorten_HR}
\end{figure}

\begin{figure}[]
\epsscale{1.0} \plotone{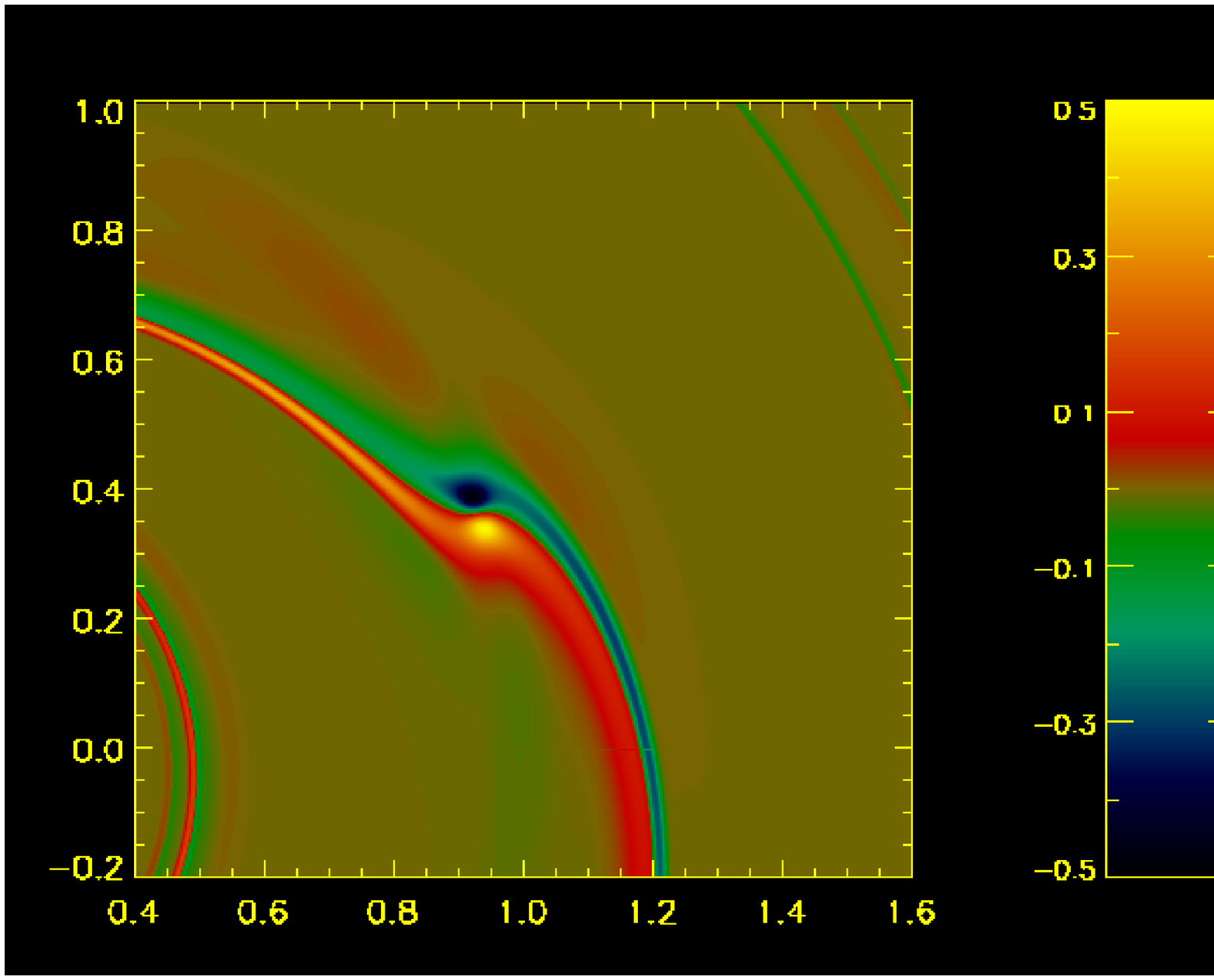}
\caption{Distribution of azimuthal density gradient at the same time
as that of Figure \ref{fig:vorten_HR}. Strong density gradient is observed
within the Roche lobe and coorbital region.
} \label{fig:drho_HR}
\end{figure}

To understand the mechanism that generates vortensity (PV) maxima within the coorbital region
in our simulations,
we note that our disk models have a non-barotropic EOS
with a radial variation of sound speed $c_s$,
which also brings in misalignment of pressure gradient and density gradient
wherever azimuthal density gradient appears ($\nabla c_s \times \nabla \Sigma \ne 0$).
This misalignment acts as a source term for the generation of vorticity
within the coorbital region,
where no shock but strong azimuthal density gradient presents.
Therefore, the large density depression in the same region naturally 
gives birth to vortensity increment.
According to  \citet{LL05}, lower resolution may result in unphysical generation of vortensity.
To further test that if our results are resolution-dependent,
we carried out a run with resolution $800\times3200$ for $\sim 70$ orbits
(due to the limitation of computational resources).
Figure \ref{fig:vorten_conv} shows the vortensity profile around coorbital regions 
for runs with different resolutions at $t \approx 70$ orbits.
They all exhibit vortensity maxima within the coorbital region,
which indicates that our results are resolution-independent.
As a further evidence that vortensity is generated around where azimuthal density gradient exists,
Figure \ref{fig:vorten_HR} and \ref{fig:drho_HR} illustrate the distribution of 
vortensity increment and azimuthal density gradient, respectively,
for the run with resolution $800 \times3200$ at $ t \sim 70$ orbits.
For a better view, they are zoomed in the neighborhood of the planet.
It is observed that vortensity is generated within the Roche lobe,
where strong azimuthal density gradient, instead of shocks, exists.
Such a generating mechanism of vortensity inside a planet's Roche lobe
has the same origin as baroclinic instability discussed in \citet{klahr03},
which contributes to global turbulence within the disk.
On the other hand, we also note that the oscillations of the torque acted on a neptune mass planet
appear much earlier in our simulations ($\sim$ 50 orbits in our Figure 6)
than they do in \citet[][$\sim$ 200 orbits in their Figure 6]{LL05},
which suggests that PV (vortensity) generation
is even more effective and common in non-barotropic disk models.
 
As show in Figure \ref{fig:drho_HR},
the azimuthal density gradient flips the sign across the planet;
thus, we expect the vortensity increment also flip the sign there.
However, 
the vortensity change within the Roche-lobe shown in Figure \ref{fig:vorten_HR}
 is always positive.
We don't fully understand the underlying physics of this phenomenon,
but suspect that this is possiblely due to vortensity mixture along
librating stream lines in the horse region,
i.e., where fluid elements make the U-turn.
A detailed and vigrous analysis of fluid motion and vortensity mixture around 
a Neptune-mass planet embedded in a non-barotropic disk
is beyond the scope of this paper,
and will be studied in a separate investigation.



Since the generation of vortensity in non-barotropic disks turns out to be quite universal,
the second instability (RWIs) triggered around vortensity extrema is probably more common 
in realistic situations than what has been thought previously;
it will have strong influence on the migration of 
a neptune mass planet in an inviscid disk.
Next, we will discuss the case in which the planet is allowed 
to move freely under the torque action from the disk.

\subsection{Simulation for a freely moving Neptune-mass planet}

In order to assess possible roles of RWIs on the migration of a planet,
we carried out two simulations in which a Neptune mass planet 
is allowed to move under the influence of
the gravitational torque from the disk.
The resolutions are $400 \times 800$ and $400 \times 1600$.
In these two simulations, the planet is fixed on the initial circular orbit 
for the first 50 orbits (slightly longer than the libration period for the 
neptune mass planet) to allow the disk to respond properly,
it is then released to move freely under the influence of the torque from the disk.
The results from these two simulations are similar to each other,
which suggests that convergence is achieved.
We use the higher resolution run to present the evolution of the system.

\begin{figure}[]
\epsscale{1.0} \plotone{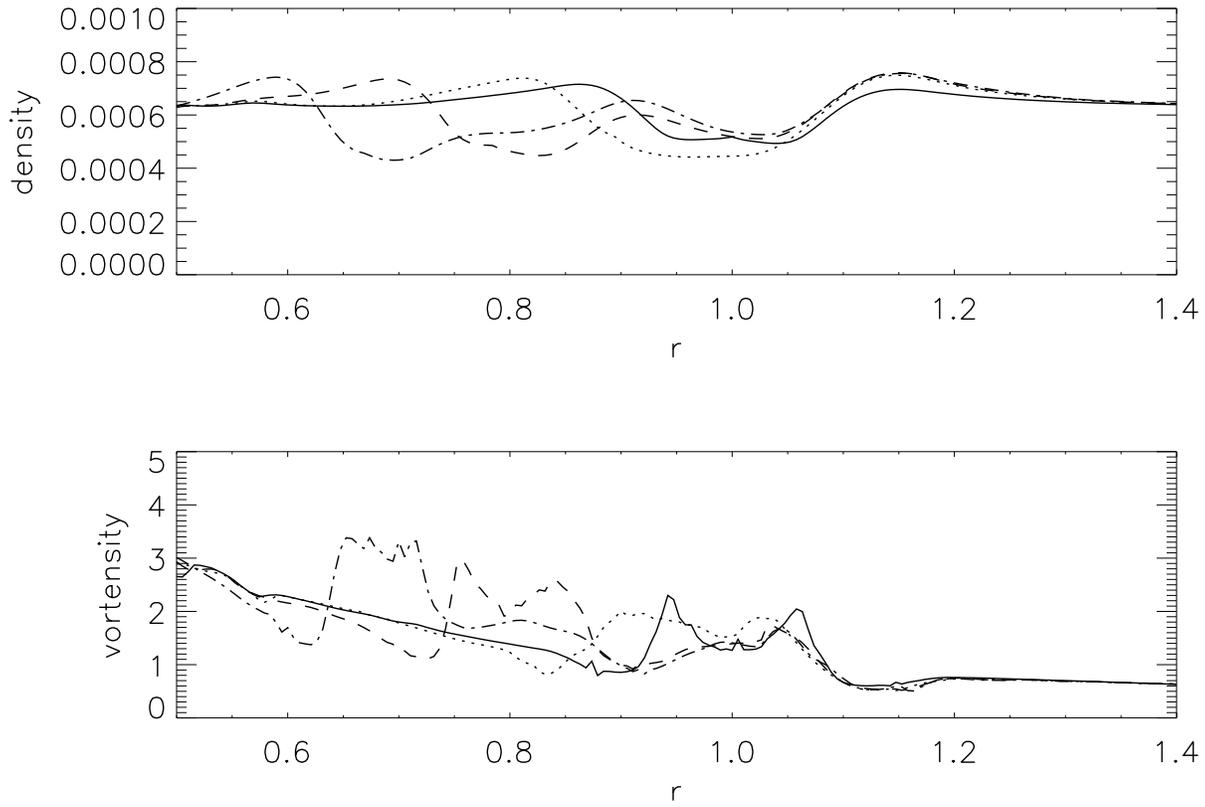}
\caption{The same as Fig. \ref{fig:rho_vort_nep} but for a run at
$t \approx 50$ (solid line), $100$ (dotted line), $150$ (dashed
line), and $200$ (dash-dotted line) orbits, in which a Neptune mass planet is allowed to move after 50 orbits.
\label{fig:rho_vort_nep_move} }
\end{figure}

Figure \ref{fig:rho_vort_nep_move} illustrates radial profiles of $\Sigma$ and 
$\xi$ averaged over the azimuthal direction at $t \approx 50, 100, 150, 200$ orbits.
It is observed that at $t=50$ orbits
the planet has created a shallow gap within the disk around $r=1$, 
the density level is enhanced at both the inner and outer edge of the gap 
($r\approx 0.88$ and $r\approx 1.12$).
Then, the planet migrates inward when $t>50$ orbits as an effect of an
overall negative total torque. As it moves in, the surface density
gap and vortensity valleys also migrate with it. On the other
hand, a shallower density gap and a vortensity valley are still preserved around its
initial orbital location. 

\begin{figure}[]
\epsscale{1.0} \plotone{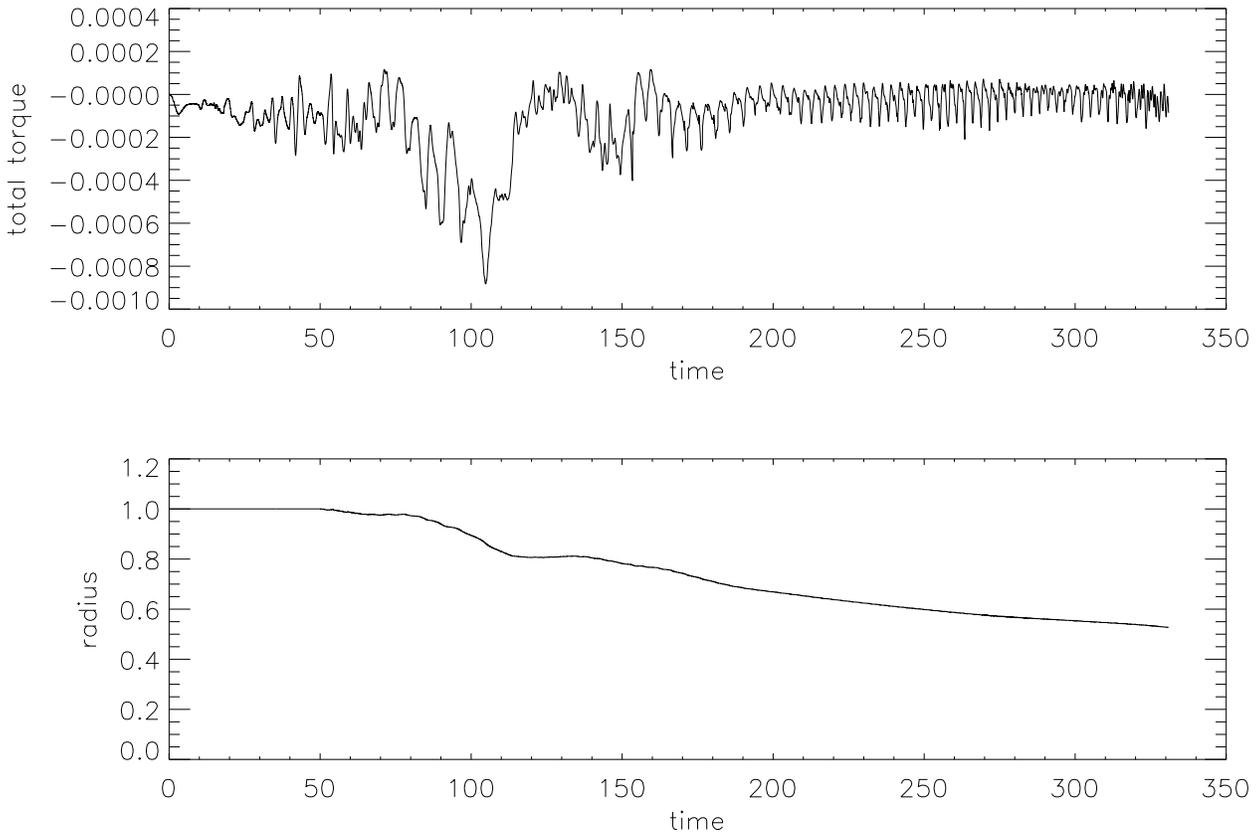}
\caption{Time evolutions of the total torque (top) exerted on a Neptune mass planet
by the disk and the radius of the planet (bottom) for the run
in which the planet is allowed to move. Raw data is shown.
A non-monotonic behavior for the migration speed is observed.
\label{fig:torq_nep_move} }
\end{figure}

Figure \ref{fig:torq_nep_move} exhibits the time evolution of the total torque acted
on the planet and its radius to the central star.
The overall torque exerted on the planet is negative in average throughout the simulation,
but undergoes some changes.
After the whole system settles down in the first 50 orbits,
large amplitude oscillations with certain period show up; 
the torque becomes increasingly negative when the planet starts to migrate inward;
it then reaches its peak in magnitude at $t \approx 100$ orbits 
($\sim 50$ orbits after releasing);
after that, it is weakened and then returns back to a very small negative value 
with smaller oscillations that have a different period from that of earlier oscillations.

\begin{figure}[]
\epsscale{1.0} \plotone{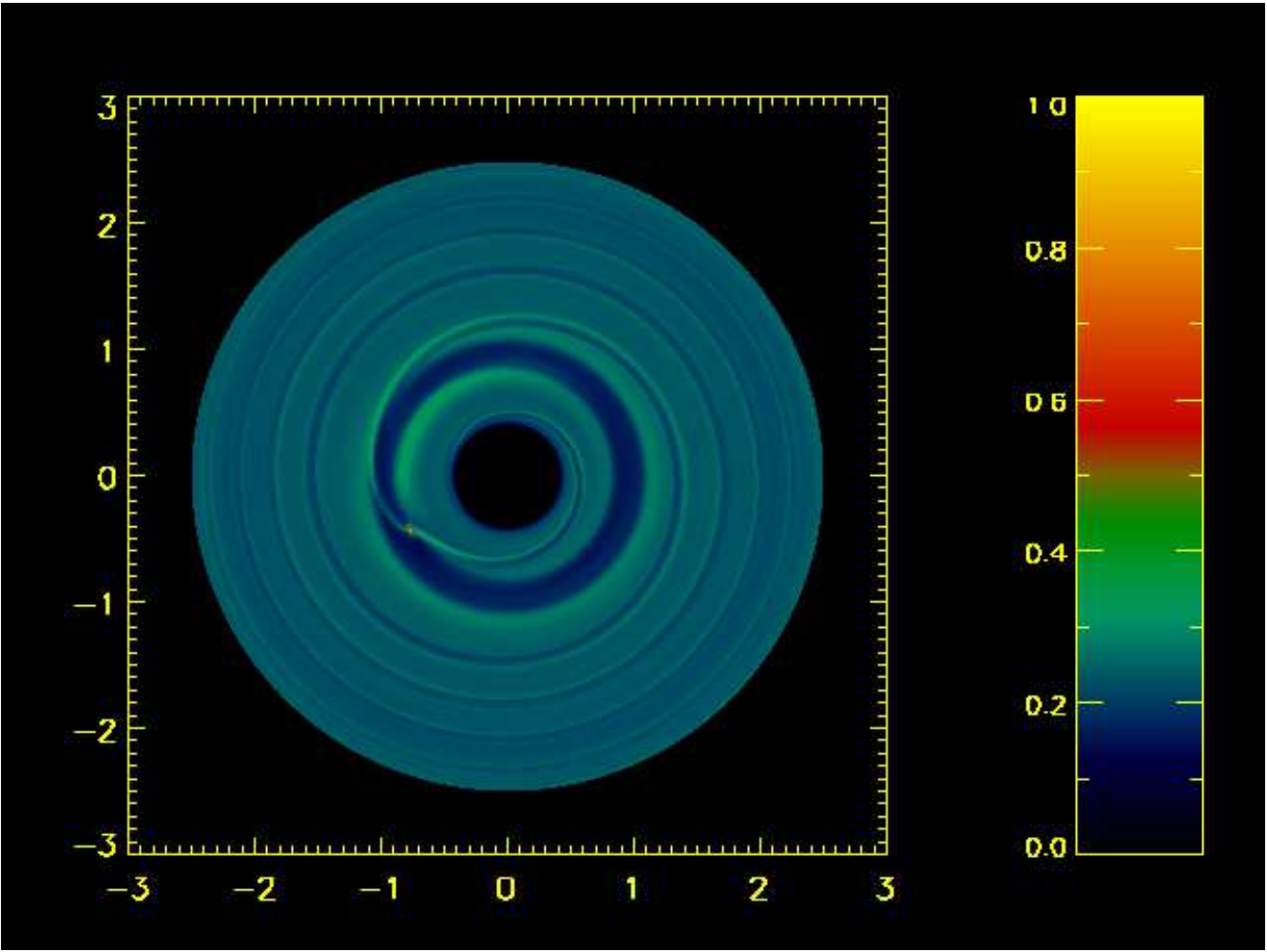} 
\caption{The same
as Fig. \ref{fig:isoden_nep} but at $t = 103$ orbits for a
run, in which a Neptune mass planet is allowed to move.
The planet is penetrating the high density region around  $r \sim 0.88$
and a fairly large density blob forms (green arc-like structure
at the edge of inner disk).
\label{fig:isoden_nep_move1} }
\end{figure}
\begin{figure}[]
\epsscale{1.0} \plotone{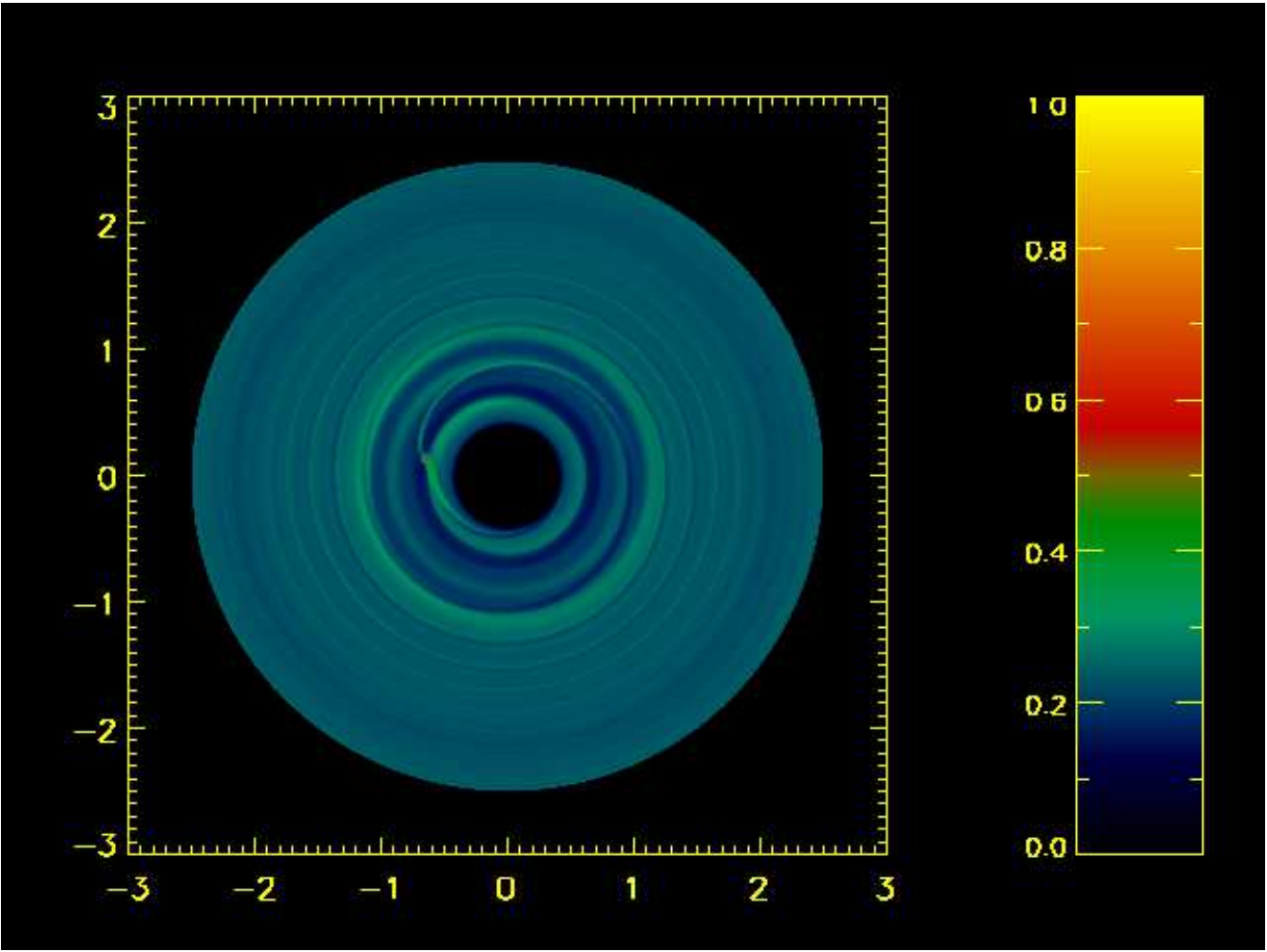}
\caption{The same
as Fig. \ref{fig:isoden_nep_move1} but at $t = 202$ orbits.
The planet has passed the original high density region around $r \sim 0.88$.
Previous large density blob wanes, smaller density blobs survive 
at several locations.
\label{fig:isoden_nep_move2} }
\end{figure}
\begin{figure}[]
\epsscale{1.0} \plotone{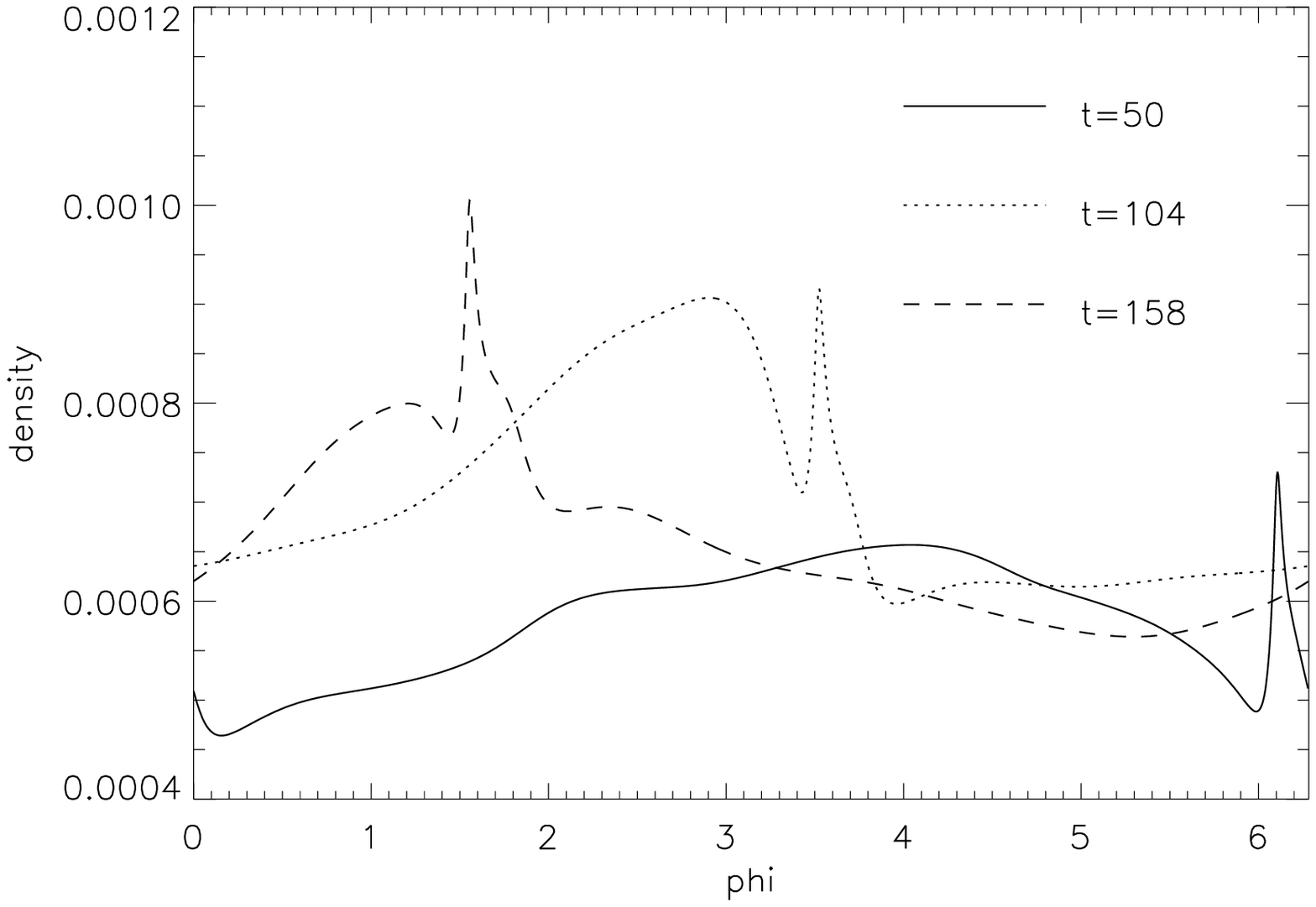}
\caption{The radially-averaged azimuthal density distribution of the ring belt
between the planet and the outer edge of the inner disk from a run
in which a neptune mass planet is allowed to move.
Note that the density blob has the highest value at $\phi\approx 3$ and $t\approx104$ when the planet
is penetrating the high density region around $r\approx 0.88$.
(The spikes reveal the locations of the planet.)
\label{fig:rhophi_nep_move2} }
\end{figure}

We carefully analyzed the evolution between $t= 80 \sim 120 $ orbits
and found that the sudden change of the torque's magnitude 
is due to the development of a relatively large density blob
between the planet and the edge of inner disk as a result of RWIs.
During this time, the density blob accumulates materials and augments very quickly
because the planet is penetrating the high density region
near $ r \sim 0.88$ (which is introduced by the planet's potential during the first
$50$ orbits, see the density profile at $t=50$ orbits).
Consequently, the density blob exerts a stronger torque onto the planet
causing the oscillation amplitude of the total torque to increase,
the averaged torque also becomes increasingly negative during $t=80 \sim 100$ orbits.
After the planet passed the high density region at $t \approx 100$ orbits,
the large density blob saturates and wanes;
the torque gradually returns back to a value close to zero.
However, smaller density blobs do survive and contribute to smaller oscillations
of the total torque in the late evolution.
Figure \ref{fig:isoden_nep_move1} shows the density colormaps of the disk at $t\approx 100$ orbits
when the planet is penetrating the high density region,
a pronounced density blob (green arc-like structure) in the edge of the inner disk is clearly visible.
For comparison, similar plot for $t \approx 200$ orbits is shown in Figure \ref{fig:isoden_nep_move2},
where previous large density blob diminishes and only small density blob survives
both in the inner disk and outer disk. 
In order to have a better view of the time behavior of the density azimuthal asymmetry 
between the planet and the outer edge of the inner disk,
the radially-averaged density azimuthal distributions of this ring-belt
at different times are ploted in Figure \ref{fig:rhophi_nep_move2}.
Besides some sharp spikes that signal the locations of the planet,
density azimuthal asymmetry with different degrees is clearly seen.
We note that the density blob has the highest value at $t\approx 104$ orbits,
which coincides with the time when the torque has the negative maximum value
and when the planet is penetrating the high density region at $r\approx0.88$. 
This further supports our analysis on the correlation between the torque magnitude
and the augmentation of the density blob.

As a consequence of this interesting phenomenon, 
the planet's radius decays quickly during $t = 80 \sim 120 $ orbits,
but the decaying slows down after $t \approx 110$ orbits
(see bottom panel of Figure \ref{fig:torq_nep_move} for this non-monotonic behavior).
We measured the migration timescale $\tau_{mig}$,
which is defined as the time needed for the planet to migrate 
from $r=1$ to the central star given a drifting speed
measured at the end of our simulation.
Our measurement yields $\tau_{mig} \approx 1300$ orbits,
which is almost an order of magnitude lower than 
that of type II migration from a similar study 
 \citep[][see their Table 1.
Note that their planet also locates at $r=1$ initially.]{NPMK00}.
This timescale is consistent with type I migration timescale,
which ought to be one or two orders of magnitude lower than that of type II
\citep{War97}.

We note that there is a trend for the averaged torque to approach zero 
at late times, this is probably because the mass of the disk interacting
with the planet decreases as it moves in.
On the other hand, the oscillation amplitude of the torque induced by
RWIs remains relatively constant, it is possible that 
as the planet moves closer to the host star
a turning point may eventually appear, where the RWI-induced torque
beats over the Lindblad torque.
If this is true, the planet may be halted at certain radial range.
Since the planet is very close to the inner edge of our computational grid
at the end of this simulation,
we decided to stop the simulation and 
will address these issue in future investigations.




In the limited time evolutions of a freely moving planet presented here, 
RWIs do not change the overall picture of inward migration;
but they have significant influence on the torque exerted on the planet
and consequently make the migration speed of the planet non-monotonic.
Further investigations are needed to study the influence of other factor
that may potentially bring in significant changes,
such as the ``releasing time" of the planet, self gravity of the disk fluid, etc.
One important issue is how to properly evaluate the torque from materials near the planet,
which is embedded in the disk.
In this paper, we computed this bruteforcely in 2D.
However, many previous studies \citep{mas02,DANGELO05,LL05} used quasi-3D treatment,
which tends to reduce the torque. 
It is very likely that if proper 3D treatment is adopted,
the total torque will be reduced and the influence of RWIs will be much stronger.
This could potentially cause the migration to be reversed.

\section{Conclusion and Discussion}

We have carried out high resolution simulations on the interaction
between a protoplanetary disk and an embedded planet
with emphasis on detailed interplay between a disk and a planet
under the influence of baroclinic generation of vortensity and non-axisymmetric RWIs.
Our results are consistent with classical analysis on
the interaction between a protoplanetary disk and an embedded planet
through Lindblad torques.
We confirmed previous studies by \citet{KLL03} and \cite{LL05}
that non-axisymmetric RWI is likely to develop under certain circumstances
and have an important influence on the migration of a planet
inside an inviscid disk.
We also found that the generation of vortensity (PV)
is more common and effective in disks with non-barotropic EOS
through the baroclinic instability,
which further favors the development of RWIs.
As the asymmetry of the density distribution induced by RWIs becomes prominent,
the resulting density blobs exert periodical and enhanced gravitational pull onto the planet
as they pass by the vicinity of the planet,
which causes the total torque received by the planet undergo large amplitude oscillations.
Although in our current simulations,
RWIs did not change the overall picture of an inward migration,
they definitely have very important and interesting effect on the migration speed.
As a side effect of an inwardly migrating planet,
RWIs introduce nonaxisymmetric density blobs along the way.
These enhanced density blobs with strong vortices may help rapid formation 
of new planet cores within them in a way suggested in \citep{klahr06, petersen06},
especially in inner regions of circumstellar disks
where rapid precipitation and coagulation of solid materials
are likely to happen \citep{sil06}. 
If these new-born cores could survive 
during the migration of a giant planet \citep{cham06,ray06},
they may produce Earth-like planets in the Habitable Zones \citep{Ji06}
or Hot Earths interior to a close-in giant planet \citep{ray06}.

The study presented here is just a very limited step among many
efforts toward understanding disk-planet interactions.
Compared to realistic situations, there are so many physics
missing in our much-simplified simulations,
the influence of RWIs
on planet migration is still not clear in cases where all the
physics are correctly taken into account. For example, if the
self-gravity of the disk ignored in this study is included in the
dynamics of the fluid, it is likely to enhance the RWIs because
the gravitational potential field is made more asymmetric by the
non-axisymmetric density distribution and more masses will be
trapped into potential wells of these density blobs. Enhanced
RWIs will exert a stronger positive torque onto the planet and 
may greatly reduce the migration speed.
On the other hand, because RWI develops mostly around the coorbital region
where corotation torques are originated,
it may have important effect on type III migration.
These issues need to be addressed in future investigations.

\acknowledgments

We thank Hui Li for useful discussions and suggestions during the course of this study.
We also thank Joel Tohline for helpful comments and
kind permission to use LSU astrophysical hydrodynamics code.
We appreciate an anonymous referee for suggesting 
various ways to analyze our results.
S.O. was partially supported by NSF grant AST-0407070 and Center for Computaion
and Technology at Louisiana State University. J.H.J. is much
grateful to Paul Butler for the hospitality during the visit stay
at DTM and also appreciates the astronomy group therein. J.H.J.
acknowledges the financial support by the National Natural Science
Foundations of China (Grants 10573040, 10673006, 10203005,
10233020) and the Foundation of Minor Planets of Purple Mountain
Observatory. The computations were performed on the Supermike
cluster, the Pelican cluster, the Nemeaux cluster at LSU,
and IBM P5 clusters of Louisiana Optical Network Initiative.

\end{document}